\documentclass[epj]{svjour}
\usepackage{graphics,amsmath}
\begin{document}
\bibliographystyle{ieeetr}

%
%\title{Effect of activity-driven connectivity pattern on epidemic process}
\title{Impact of temporal connectivity patterns on epidemic process}

\author{Hyewon Kim\inst{1} 
\and Meesoon Ha\inst{2}%
\thanks{\emph{} msha@chosun.ac.kr}%
\and Hawoong Jeong\inst{1,3}% etc
\thanks{\emph{} hjeong@kaist.edu}
}                     % Do not remove
\institute{Department of Physics, Korea Advanced Institute of Science and Technology, Daejeon 34141, Korea
\and Department of Physics Education, Chosun University, Gwangju 61452, Korea
\and Institute for the BioCentury, Korea Advanced Institute of Science and Technology, Daejeon 34141, Korea}
\date{Received: date / Revised version: \today}
\abstract{To provide a comprehensive view for dynamics of and on many real-world temporal networks, we investigate the interplay of temporal connectivity patterns and spreading phenomena, in terms of the susceptible-infected-removed (SIR) model on the modified activity-driven temporal network (ADTN) with memory. In particular, we focus on how the epidemic threshold of the SIR model is affected by the heterogeneity of nodal activities and the memory strength in temporal and static regimes, respectively. While strong ties (memory) between nodes inhibit the spread of epidemic to be localized, the heterogeneity of nodal activities enhances it to be globalized initially. Since the epidemic threshold of the SIR model is very sensitive to the degree distribution of nodes in static networks, we test the SIR model on the modified ADTNs with the possible set of the activity exponents and the memory exponents that generates the same degree distributions in temporal networks. We also discuss the role of spatiotemporal scaling properties of the largest cluster and the maximum degree in the epidemic threshold. It is observed that the presence of highly active nodes enables to trigger the initial spread of epidemic in a short period of time, but it also limits its final spread to the entire network. This implies that there is the trade-off between the spreading time of epidemic and its outbreak size. Finally, we suggest the phase diagram of the SIR model on ADTNs and the optimal condition for the spread of epidemic under the circumstances.
%Many real-world temporal networks, such as social and biological networks, exhibit non-trivial connectivity patterns in their interactions. Here, we investigate the effects of connectivity patterns on an epidemic process by considering the susceptible-infected-removed (SIR) epidemic model in a modified activity-driven temporal network with memory. We show that heterogeneity of nodal activities reduces the epidemic threshold, while strong ties between nodes shift the threshold to larger values. In particular, the epidemic threshold is closely related to the effective size of the connections, such as the largest cluster size. We find that highly active nodes can trigger the epidemic in a short period of time, but eventually inhibit its spread to the entire network; in other words, there is a trade-off between the timing and the prevalence of the epidemic. Finally, we discuss optimal strategies for the spread of an epidemic according to the stages.
%
 % end of PACS codes
} %end of abstract

\maketitle
\section{Introduction}
\label{intro}

Epidemic processes on networks have been widely used to understand the influence of network features, such as hubs and community structures, on spreading phenomena. In the past, most studies focused on {\it static} networks, where network topologies are {\it fixed in time}. However, real-world networks are highly dynamic in time, so-called {\it temporal} networks, where both nodes and links {\it appear or disappear over time}~\cite{Holme2012,Holme2015}. The availability of high-resolution data on time-ordered interactions provides a wealth of information for the interaction patterns of nodes~\cite{Cattuto2010,Stehle2011,Barrat2013,Fournet2014}. 

Among temporal network models proposed to explain the heterogeneity of connectivity patterns, the activity-driven temporal network (ADTN) model~\cite{Perra2012}, in which each node has its own activity, was successful to generate the variety of temporal networks, distinct from static ones. In the ADTN model, at each time step, nodes are activated according to pre-assigned activities, and the active nodes generate links to randomly selected nodes. Since in real networks, nodal connections are not completely random but rather the results of a non-Markovian process with memory~\cite{Granovetter1973,Onnela2007,Miritello2011,Vestergaard2014,Saramaki2014,Rocha2016,Kobayashi2019}, the effect of memory was also widely considered as well as attributes: social ties, burstiness, and communities~\cite{Karsai2014,Medus2014,Kim2015,Sun2015,Ubaldi2016,Kim2018}. Memory induces heterogeneous temporal connectivity patterns with non-trivial correlations.
Based on the earlier studies of epidemic processes~\cite{Rocha2011,Karsai2011,Castellano2012,Scholtes2014,Rosvall2014,Gleeson2016}, non-trivial connectivity patterns enable to either speed up or slow down the spread of epidemic. Such impacts of connectivity patterns were studied on a variety of temporal networks~\cite{Rizzo2014,Liu2014,Tizzani2018,Nadini2018,Valdano2018,Moinet2018,Williams2018,Petri2018,Li2019},
which mostly focused on the effect of non-Markovian dynamics on the spread of epidemic, and discussed changes in outcomes.

Regarding the contrasting effects of strong ties, Sun and coworkers~\cite{Sun2015} studied the role of memory in epidemic processes on ADTN models with the comparison of empirical data analyses, where memory can decrease or increase the epidemic threshold. Most recently, Tizzani and coworkers~\cite{Tizzani2018} has extended the study with the elaborated memory that is tunable as the aging effect. Based on the activity-based mean-field (ABMF) approximation as well as numerical simulations, they claimed that memory are related to the reinforcement of activity-fluctuation effects. As a result, the epidemic threshold can be either decreased or increased, compared to the ABMF asymptotic value depending on the initial setup. However, there is still a lack of comprehensive understanding of the interplay between the epidemic outcomes and the topological features of ADTNs with memory.    

In this paper, we investigate the role of temporal connectivity patterns in spreading phenomena as well as the optimal condition for the spread of epidemic, in terms of the susceptible-infected-recovered (SIR) model on the modified ADTN model~\cite{Kim2015} with memory. In particular, we focus on how temporal connectivity patterns affect the epidemic threshold and the outbreak size in the SIR model. We find that it is crucial to consider the effective size of temporal networks, such as the spatiotemporal properties of the largest cluster size and the maximum degree. In the presence of memory, the spread of epidemic becomes slowing down and the epidemic threshold gets larger than the memoryless case. The heterogeneity of nodal activities may enhance the spread of epidemic at the initial spreading time of the epidemic, but may inhibit it later due to no reinfection. The trade-off exists between the spreading time and the outbreak size, which depends on the topological condition to reach its best prevalence.

The rest of the paper is organized as follows. In Sec.~\ref{networkpart}, we briefly describe the modified ADTN model with memory, and show time-varying network topologies. In Sec.~\ref{epidemicpart}, the SIR model is tested on the modified ADTN model. Based on numerical results, we propose the phase diagram of the SIR model on ADTNs. We also discuss the optimal condition to achieve both the early onset and wide prevalence of the epidemic under the circumstances. Finally, we summarize our results and provide some remarks in Sec.~\ref{summary}.

\begin{figure}[]
\center
\resizebox{0.475\textwidth}{!}{%
\includegraphics{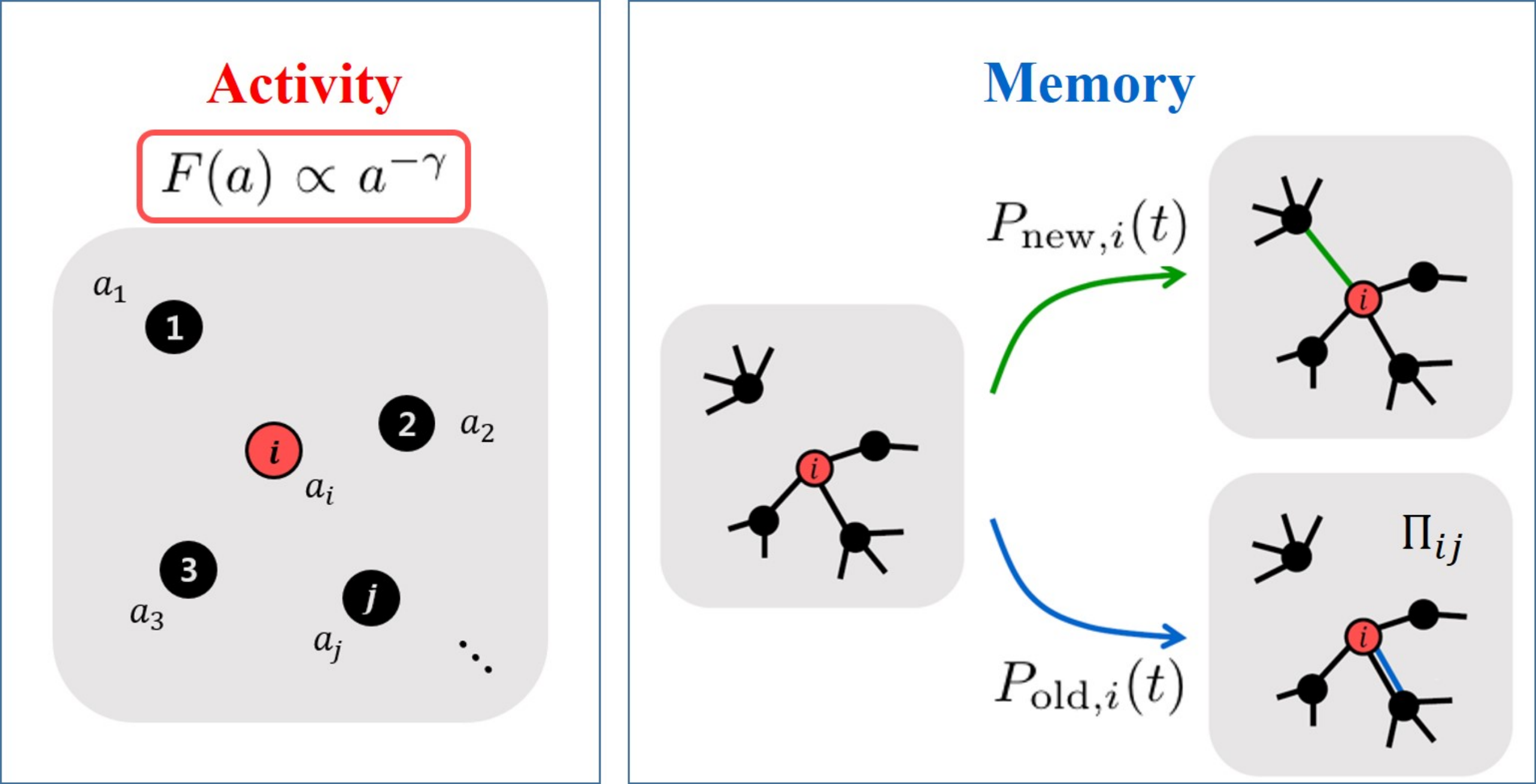}
}
\caption{The modified ADTN model dynamics was illustrated as two panels: The left one represents how to activate a node, while the right one represents how for the activated node to be connected to another node. Here $F(a)$ is the distribution of nodel activitites, $P_{{\rm new (old)},i}$ is the probability that the activated node $i$ is connected to a new (old) neighbor, and $\Pi_{ij}$ is the preference probability between node $i$ and node $j$, see Eq.~\eqref{eq:model} for detailed mathematical expressions. The background links shown in the memory panel, correspond to the time-aggregated version, which are not shown at time $t$.}
\label{fig1:model}
\end{figure}
\begin{figure*}[]
\center
\resizebox{0.95\textwidth}{!}{%
\includegraphics{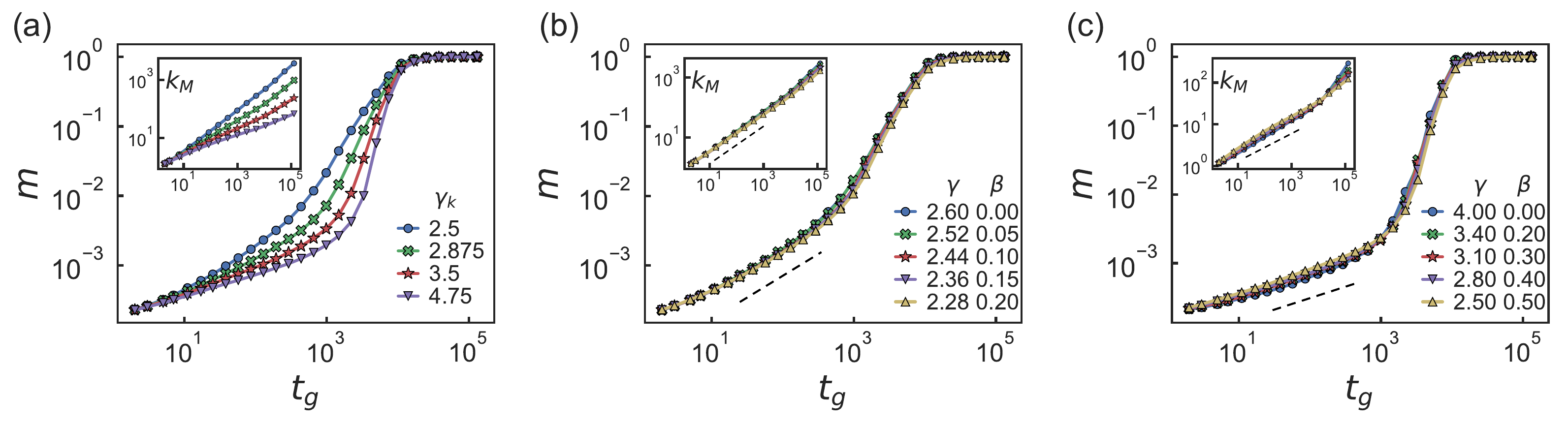}
}
\caption{The fraction of the largest cluster (giant connected component, GCC), $m(t_g)=M(t_g)/N$, is plotted as a function of the \textit{growth} time, $t_g$ in the modified ADTN model, where the total number of nodes is $N=10^4$: (a) $\beta=0, 0.2, 0.4,$ and $0.6$ (from top to bottom) with $\gamma=2.5$. From $\gamma_k=(\gamma-\beta)/(1-\beta)$, (b) $\gamma_k=2.60$ and (c) $\gamma_k=4.00$, respectively. The insets show the maximum degree, $k_M$, as a function of $t_g$. The dashed lines are guide for the eye, whose slopes represent the values of  the growth exponent $\eta=1/(\gamma_k-1)$, so (a) $\eta=5/8$ and (b) $\eta=1/3$. Numerical data are averaged over $500$ network samples.}
\label{fig2:GCC}
\end{figure*}

\section{Modified ADTN model}
\label{networkpart}

The modified activity-driven temporal network (ADTN) model~\cite{Kim2015} is a simple variant of the ADTN model~\cite{Perra2012}, which allows to control the memory strength with the memory exponent $\beta$, provided that the heterogeneity of nodal activities is controlled by the activity exponent $\gamma$ in the original ADTN model. In this section, we briefly describe how to generate an ADTN with memory and discuss  relevant physical properties, compared to the static case.

\subsection{How to generate an ADTN with memory}
\label{ADNM}

We start with $N$ disconnected nodes, which have the pre-assigned own activities, $a_i$ ($i\in \mathcal{N}=\{1, ...,N\}$) from the activity distribution $F(a)$, where $i$ is the nodal index. For the heterogeneity of nodal activities, we consider a power-law activity distribution with the activity exponent $\gamma$, so that $F(a)\sim a^{-\gamma}$. 

As shown in Fig.~\ref{fig1:model}, at each time step $t$, we choose node $i$ to be activated with the probability $p_i=a_i/a_{\text{max}}$, where $a_{\text{max}}=\max_{j=1}^{N}\{a_j\}$. The activated node $i$ can generate either a new link to a node among the nodes that have never had a link to node $i$ or an old one among the nodes that have ever had a link to node $i$ until time $t$. For the former, with probability $P_{{\rm new},i}(t)$, node $i$ connects to a randomly selected one among new nodes. For the latter, with probability $P_{{\rm old},i}(t)=1-P_{{\rm new},i}(t)$, it connects to an old node $j(\ne i)$ 
with the preference probability $\Pi_{ij}(t)$.

We define $P_{{\rm new},i}(t)$ and $\Pi_{ij}$ as follows: 
\begin{align}
P_{{\rm new},i}(t) = As_i(t)^{-\beta}~~ \mbox{and} ~ ~\Pi_{ij}(t) =& \frac{w_{ij}(t)}{\sum_{l\in\mathcal{N}}w_{il}(t)},
\label{eq:model}
\end{align}
where $\beta$, $s_i(t)$, and $w_{ij}(t)$ are the memory exponent from 0 (memoryless) to 1, the accumulated strength of node $i$, and the accumulated link-weight between nodes $i$ and $j$ up to time $t$, respectively. For the simplicity, we set $A=1$ without loss of generality. For the case of $\beta=0$ (memoryless), $P_{{\rm new},i}(t)$ becomes time-independent, while for the case of $\beta \neq 0$, the activated node $i$ prefers to choose an old link rather than to make a new one. Thus, the larger $\beta$, the stronger tie (the larger link-weight) between already connected nodes. In the time-accumulated network representation of the modified ADTN model with $\gamma$ and $\beta$, the generated temporal network becomes the weighted scale-free network that has the following statistics of network properties: $P(s) \sim s^{-\gamma_s}$ with $\gamma_s=\gamma$, $P(k) \sim k^{-\gamma_k}$ with $\gamma_k=\frac{(\gamma-\beta)}{(1-\beta)}$, and $P(w) \sim w^{-\gamma_w}$ with $\gamma_w=\gamma$ as $\beta \rightarrow 1$ (see the detailed derivation can be found in Ref.~\cite{Kim2018}).

\subsection{Growth of largest cluster and maximum degree}
\label{ADNMtopology}

The largest cluster size, $M$, and the maximum degree $k_M$, (\textit{i.e.}, the effective size of connections) are important features that compose the network properties of a ADTN. In order to reveal the effects of connectivity patterns on time-varying topologies, we  revisit the growth pattern of the network structure discussed in our previous study~\cite{Kim2018} for dynamic topologies as a function of sequence $t$, defined as the number of the events (links). 

In Fig.~\ref{fig2:GCC}, the main plots show the growth patterns of the fraction of the largest cluster size (giant connect component, GCC), $m=M/N$, up to the accumulated time $t_g$, \textit{i.e.} for various $\gamma$ and $\beta$ against $t_g$. There are three regimes in the growth of GCC: In the dynamic regime, it grows depending on connectivity patterns of the corresponding network properties. In the intermediate-time regime, the finite-size effect comes in, and, it eventually saturates to $1$ in the static regime.  Based on the dynamic scaling of the GCC in our early study~\cite{Kim2018}, we readdress the growth pattern of the GCC in the dynamic regime is related to the nature of the temporal network, and plays a crucial role in network analyses. In Fig.~\ref{fig2:GCC} (a), we test the role of $\beta$ in the growth of the GCC at $\gamma=2.5$, which represents $\gamma_k=2.5~(\beta=0),~2.875~(0.2), 3.5~(0.4),~\mbox{and}~4.75~(0.6)$. As $\beta$ (the strength of memory) increases, the GCC grows slowly because nodes prefer to contact with already connected (old) nodes. For this case, the growth pattern of the GCC depends on the degree distribution obtained from temporal connectivity patterns between nodes.

As shown in (b) and (c) of Fig.~\ref{fig2:GCC}, the temporal growth pattern of the GCC can be categorized by the degree exponent $\gamma_k=(\gamma-\beta)/(1-\beta)$ among different connectivity patterns with a variety of settings with $\gamma$ and $\beta$. 
In the dynamic regime, the GCC exhibits the same scaling behavior as $M\sim k_M$, so that
\begin{align}
M(t)\sim k_M(t)\sim t^\eta
\end{align}
with the natural cutoff degree scaling $\eta=1/(\gamma_k-1)$. 

For the small values of $\gamma$, the GCC is created by a few highly active nodes, resulting in $M(t)$ growing quickly, while for the large values of $\beta$, the growth of the GCC is delayed due to the establishment of strong ties related to memory. As a result, the growth pattern of the GCC for small $\gamma$ and large $\beta$ is effectively the same as that for large $\gamma$ and small $\beta$. The growth of the maximum degree in the time-accumulated network is shown in the insets of Fig.~\ref{fig2:GCC}, which implies that the effective size of nodal connections with different activities and memory strengths can be the same if networks have the same degree exponent. 

Since temporal connectivity patterns may affect not only network topologies but also dynamic processes on them, it is interesting to discuss the effect of temporal connectivity patterns on spreading phenomena. In the next section, we revisit the susceptible-infected-recovered (SIR) model on temporal networks~\cite{Sun2015,Tizzani2018}, in terms of ADTNs with memory, and provide a comprehensive view for the epidemic threshold in temporal networks, compared to that in static one.
\begin{figure*}[]
\center
\resizebox{0.95\textwidth}{!}{%
\includegraphics{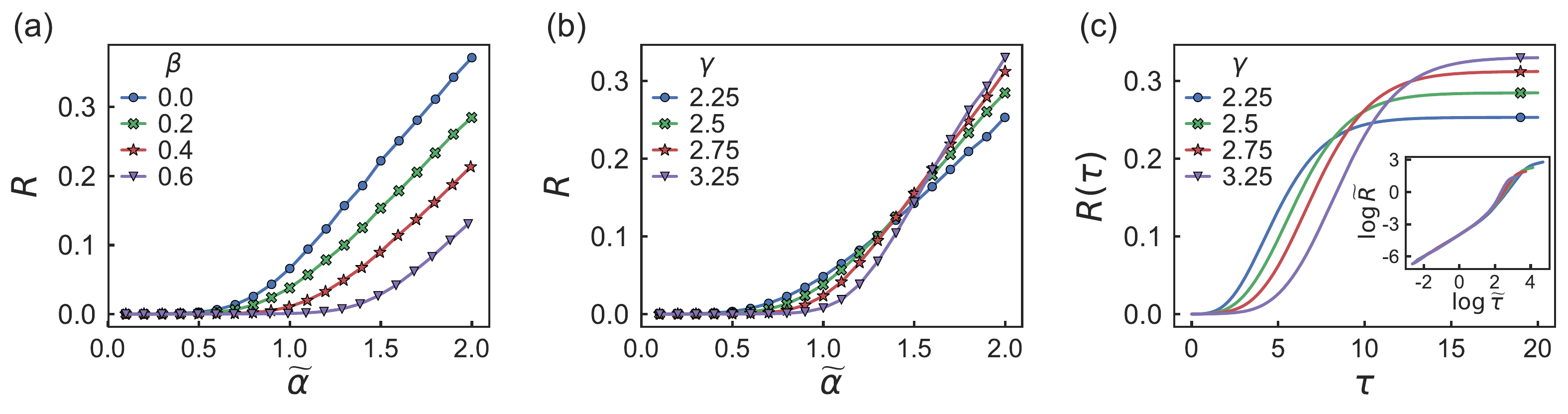}
}
\caption{In the steady-state limit, the fraction of final recovered nodes (outbreak size), $R$, is plotted as a function of $\widetilde{\alpha}$: (a) the memory effect is tested on the SIR model with $\gamma=2.5$ for $\beta=0, 0.2, 0.4,$ and $0.6$, and (b) the role of the activity heterogeneity is tested in the SIR model with $\beta=0.2$ for $\gamma=2.25, 2.5, 2.75,$ and $3.25$ in the presence of memory. (c) The temporal behavior of $R$ at $\widetilde{\alpha}\approx 2.0$ of (b) is drawn against $\tau=t/N$, where the inset shows the best collapse until the first kink part of the each curve from the scaling ansatz from Eq.~\eqref{eq:scaling}, where $\widetilde{R}=R k_M$, $\widetilde{\tau}=\tau k_M$, and $k_M(t)\sim t^{\eta}$ with $\eta=1/(\gamma_k -1)$. Here $N=10^4$ and numerical data are averaged over 100 network samples with 500 samples of different initial conditions.}
\label{fig3:b+r+temporal}
\end{figure*}

\section{Epidemics on ADTN with memory}
\label{epidemicpart}

\subsection{SIR dynamics}
\label{epidemicmodel}

The modified ADTN in which an epidemic spreads can be represented as a set of time-ordered subnetworks $G_n$ ($n=1, 2, ...$).  Each subnetwork $G_n$ represents a network accumulated during the interval $[(n-1)\Delta t, n\Delta t)$. In this paper, we set $\Delta t=10$ to investigate the effect of temporal connectivity patterns on epidemics. 

We consider the classic SIR model, where a node can be one of three states: Susceptible ({\bf S}) nodes are uninfected (healthy) that can be infected at a rate of $\alpha$ if they are in contact with infected neighboring nodes at time $t$. Infected ({\bf I}) nodes spread the epidemic to susceptible nodes that also recover themselves at a rate of $\mu$. Then I nodes become recovered ({\bf R}) ones are permanently immune that are not further involved in dynamic. In this sense, {\bf R} nodes are also often called \textit{removed} ones. The dynamics of the SIR model is as follows:
\begin{align}
{\rm \bf S} + {\rm \bf I}& \xrightarrow{~\alpha~} 2{\rm \bf I},\nonumber\\
 ~{\rm \bf I}& \xrightarrow{~\mu~} {\rm \bf R}.
\end{align}

In the SIR model,  the initial states of all nodes are {\bf S}, except for one seed {\bf I} node that is selected at random. At each network step $n$, the epidemic spreads in the network $G_n$. Each {\bf S} node in $G_n$ is infected with probability $p_\alpha$ if it contacts with an {\bf I} node. At the same time, all {\bf I} nodes change to {\bf R} nodes with probability $p_\mu$. At the next network step $n+1$, the network $G_n$ is changed to $G_{n+1}$, and the dynamic process is repeated in $G_{n+1}$. This procedure repeats until the last subnetwork. Note that $p_\alpha$ is the probability per contact, so $p_\alpha=\alpha/\langle k\rangle$,
where $\langle k\rangle$ is the average degree per unit step. Here we use $N=10^4$ in the modified ADTN model and $\mu=10^{-3}$ in the SIR model. Numerical data are averaged over $100$ network configurations with $500$ independent simulations of initial conditions.

\subsection{Epidemic threshold and outbreak size}
\label{result1}

In the SIR model, the outbreak size and the epidemic threshold are the most interesting physical quantities. The outbreak size $R$ can be measured as the fraction of {\bf R} nodes and the epidemic threshold $\widetilde{\alpha}=\alpha/\mu$ indicates the ordinary bond percolation transition~\cite{Cohen2002}, which separates the epidemic phase from the infection-free phase. Below $\widetilde{\alpha}_c$, $R\to 0$ in the thermodynamic limit ($N\to\infty$).

\begin{figure}[]
\center
\resizebox{0.425\textwidth}{!}{%
\includegraphics{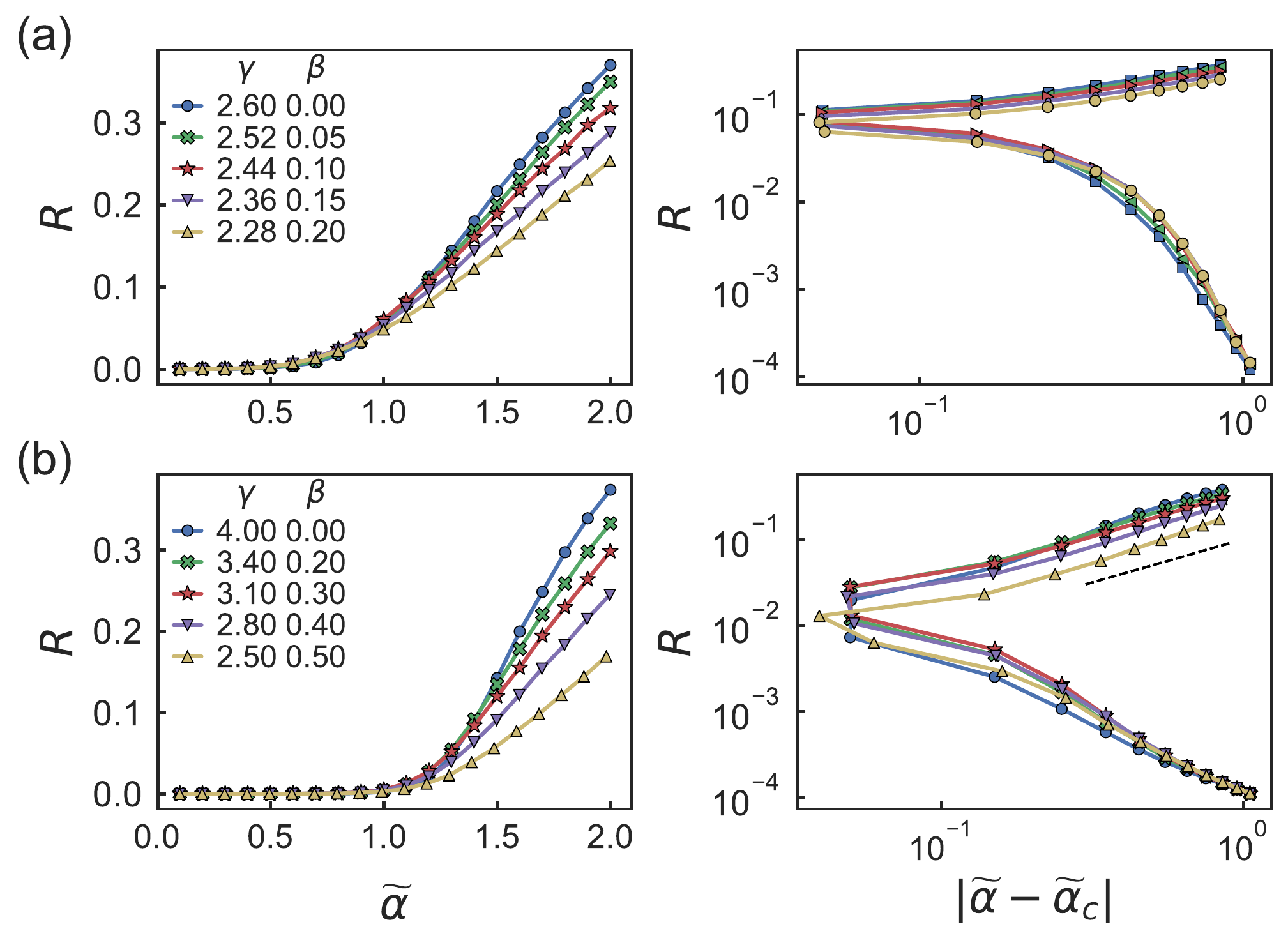}
}
\caption{For the proper combinations of $\beta$ and $\gamma$ to generate the same degree exponent $\gamma_k$, $R$ is plotted  in the left panel as the format of real scales against $\widetilde{\alpha}$ and in the right panel as the format of double-logarithmic scales against $|\widetilde{\alpha}-\widetilde{\alpha}_c|$, where $\widetilde{\alpha}_c$ depends on $\gamma_k$: (a) $\gamma_k=2.6$ and (b) $\gamma_k=4.0$, where $N=10^4$ and $\widetilde{\alpha}_c$ is numerically obtained. The dashed line in the  right panel of (b) is guide for the eye, whose slope is $\beta_{_{\rm SIR}}=1/(\gamma_k-3)$ for $\gamma_k > 3$.}
\label{fig4:test+scaling}
\end{figure}

In modified ADTNs as $\gamma$ and $\beta$ vary, we discuss how temporal connectivity patterns affect $R$ and $\widetilde{\alpha}_c$ as shown in Fig.~\ref{fig3:b+r+temporal} and Fig.~\ref{fig4:test+scaling}. In Fig.~\ref{fig3:b+r+temporal}, we test (a) the role of memory in the final outbreak and (b) the heterogeneity of nodal activities in the presence of memory, respectively. Moreover, we investigate temporal behaviors of (b) at $\widetilde{\alpha}\approx 2.0$ in the supercritical regime as (c), where we observe some trade-off between the spreading time and the outbreak size according to the activity exponent. Based on the detailed analysis, we find that the early-time growth of $R$ is governed by the growth of the GCC (or the maximum degree), so that we can collapse temporal data of $R$ by using the following scaling ansatz:
\begin{align}
R(\tau)=k_M f(\tau k_M),
\label{eq:scaling}
\end{align} 
where $\tau$ is the rescaled time by $N$, and $k_M\sim t^{\eta}$ with $\eta=1/(\gamma_k-1)=(\gamma-1)/(1-\beta)$ in the modified ADTN model~\cite{Kim2015,Kim2018}. The numerical confirmation is provided as the inset of (c) in Fig.~\ref{fig3:b+r+temporal}.
\begin{figure*}[]
\center
\resizebox{0.95\textwidth}{!}{%
\includegraphics{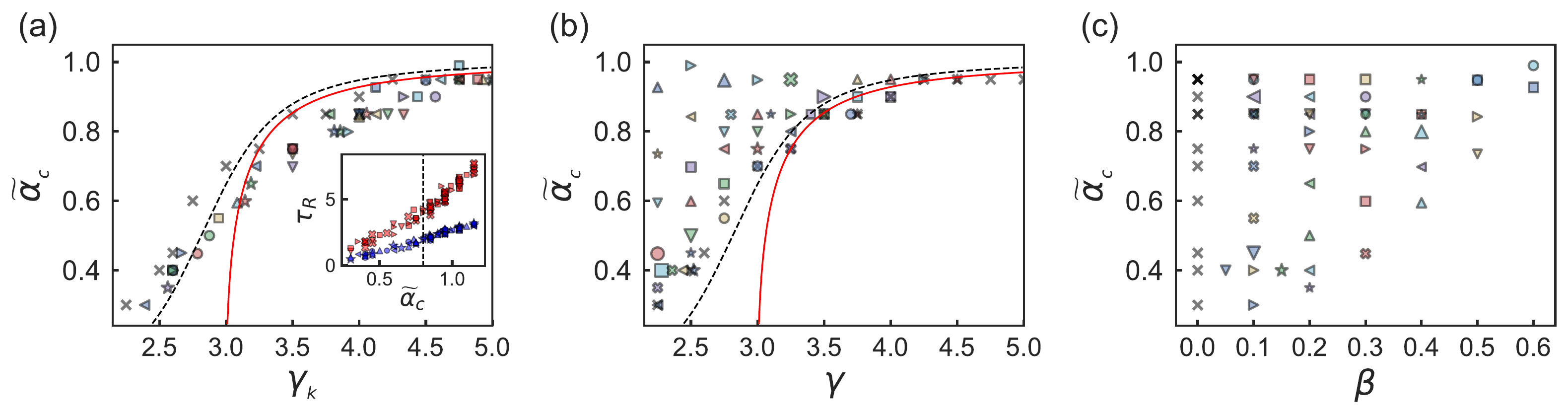}
}
\caption{The phase diagram of the SIR model is drawn on the modified ADTN model: the epidemic threshold, $\widetilde{\alpha}_c$, is plotted against (a) the degree exponent, $\gamma_k$, (b) the activity exponent, $\gamma$, and (c) the memory exponent, $\beta$. In (a),  the dashed (black) and solid (red) lines are guide for the eye, which are the ABMF results for $N=10^4$ and the thermodynamic limit, respectively (see Eq.~\eqref{eq:ABMF} for detailed analytic expressions), the inset of which represent the plots of the rescaled access time, $\tau_R$, when $R=10^{-3}~(\mbox{bottom, blue})$ and $R=10^{-2}~(\mbox{top, red})$ against $\widetilde{\alpha}_c$ of each dataset. The values of $\tau_R$ are taken from the main plot of (c) in Fig.~\ref{fig3:b+r+temporal}. The vertical dotted line of the inset in (a) corresponds to the data at $(\gamma_k,\widetilde{\alpha}_c)\approx (3.85, 0.80)$ in the main plot. Here numerical data are averaged over 100 network samples and 500 samples of different initial conditions for $N=10^4$.}
\label{fig5:PD+}
\end{figure*}
\begin{figure}[]
\center
\resizebox{0.475\textwidth}{!}{%
\includegraphics{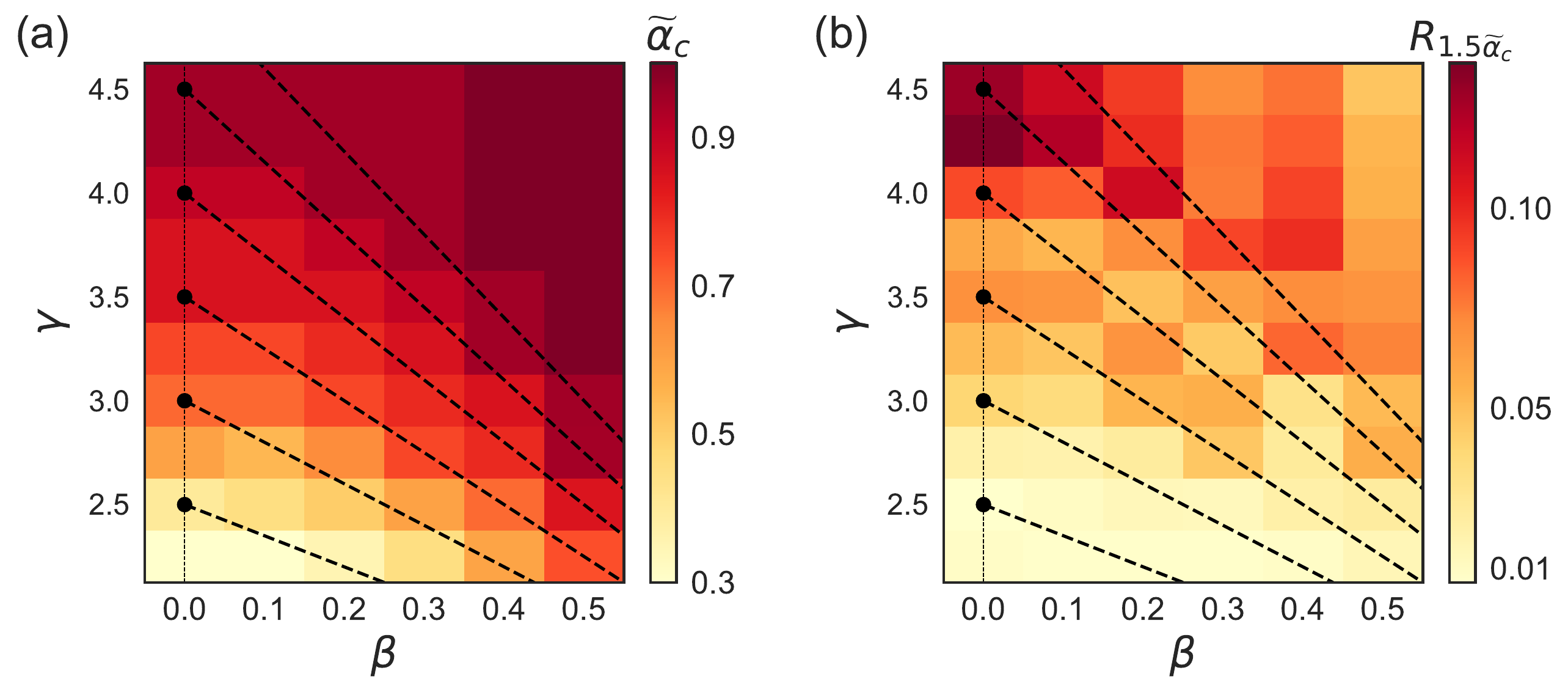}
}
\caption{The optimization conditions of the epidemic threshold, $\widetilde{\alpha}_c$, and the outbreak size, $R_{1.5\widetilde{\alpha}_c}$ at $\widetilde{\alpha}=1.5\widetilde{\alpha}_c$, of the SIR model on ADTNs with and without memory in the plane of $\gamma$ and $\beta$: (a) the heatmap of $\widetilde{\alpha}_c$ and (b) that of $R_{1.5\widetilde{\alpha}_c}$. The dashed lines are guide for the eye, which represent the same values of $\gamma_k=(\gamma-\beta)/(1-\beta)$.}
\label{fig6:heatmap}
\end{figure}

Based on numerical simulations, we find that $R$ and $\widetilde{\alpha}_c$ depend on not only temporal connectivity patterns but also network degree structures. Thus, we suggest the same analysis in static  networks for the data obtained from the same degree exponent $\gamma_k$, consisting of different $\gamma$ and $\beta$. In Fig.~\ref{fig4:test+scaling}, we test (a) $\gamma_k=2.60$ and (b) $\gamma_k=4.00$. As expected, the epidemic threshold $\widetilde{\alpha}_c$ is still the same even if temporal connectivity patterns are different, while the value of the outbreak size $R$ in the supercritical regime is memory-dependent. The strong the strength of memory (the larger $\beta$), the smaller the outbreak size. 

To numerically estimate $\alpha_c$ in finite systems, we propose the criterion for the criticality in ADTNs: At the criticality of the SIR model, we assume $R_c=cN^{-\eta}$ from Eq.~\eqref{eq:scaling}, where we set $c=10^{-1}$. Using the numerical estimation for the value of $\widetilde{\alpha}_c$ at $R=R_c$, we replot $R$ against $|\widetilde{\alpha}-\widetilde{\alpha}_c|$ in the right panels of Fig.~\ref{fig4:test+scaling}  (a) and (b), respectively. The dashed line in (a) is a guide for the eye, whose slope corresponds to $\beta_{\rm_{SIR}}=1/(\gamma_k-1)$ for $\gamma_k>3$ from $R\sim (\widetilde{\alpha}-\widetilde{\alpha}_c)^{\beta_{_{\rm SIR}}}$ in the supercritical regime.

Our numerical estimates of the epidemic threshold for various cases are shown in Fig.~\ref{fig5:PD+}, where they are is plotted against (a) $\gamma_k$, (b) $\gamma$, and $\beta$, respectively. In Fig.~\ref{fig5:PD+} (a) and (b), we also provide the analytic curve obtained from the activity-based MF (ABMF) approximation~\cite{Perra2012,Sun2015,Tizzani2018} for the memoryless case, which is as follows:
\begin{align}
\widetilde{\alpha}_{c,{\rm ABMF}}=\frac{2\langle a\rangle}{\langle a\rangle+\sqrt{\langle a^2\rangle}},
\label{eq:ABMF}
\end{align}
where $a$ is the nodal activity.

Interestingly, we find that the epidemic threshold seems to be well described by $\gamma_k$. Moreover, for $\gamma<4.0$, the memory effect becomes relevant to increase the epidemic threshold, compared to that for the memoryless case, which is consistent with the result of the most recent study~\cite{Tizzani2018}.

\subsection{Optimal conditions for the spread of epidemic}
\label{result2}

Figure~\ref{fig5:PD+} (a) shows the phase diagram of the SIR model on ADTNs with memory as $\gamma_k$ varies. For the same $\gamma$, the epidemic threshold $\widetilde{\alpha}_c$ increase as $\beta$ increases, which is shown in Fig.~\ref{fig5:PD+} (b). For the same $\beta$, it increases as $\gamma$ increases, which is shown in Fig.~\ref{fig5:PD+} (c).  Regarding the spreading time (speed) that is shown in the inset of Fig.~\ref{fig5:PD+}, the smaller the epidemic threshold, the faster the spreading speed. However, it is not always guaranteed that the outbreak size is also large.  

To discuss the optimal conditions for the spread of epidemic, we present the epidemic threshold and the outbreak size as the format of heatmap in the plane of the activity exponent ($\gamma$) and the memory exponent ($\beta$), see Fig.~\ref{fig6:heatmap}. As shown in Fig.~\ref{fig6:heatmap} (a), the smaller the strength of memory ($\beta\to 0$) and the larger the activity heterogeneity ($\gamma \to 2$), the smaller the epidemic threshold ($\widetilde{\alpha}_c\to 0$). In the supercritical regime, $\widetilde{\alpha}>\widetilde{\alpha}_c$, see Fig.~\ref{fig6:heatmap} (b),  the smaller the strength of memory ($\beta\to 0$) and the larger the heterogeneity of nodal activities ($\gamma \to 4.5$), the larger the outbreak size. This is due to the trade-off between the initial spreading speed and the final outbreak size.  

In other words, for both the early onset and the final prevalence, the memory effect in temporal connectivity patterns diminishes the efficiency of $\widetilde{\alpha}$. For the same $\beta$, the heterogeneity of nodal activities enhances the efficiency of $\widetilde{\alpha}$ for the early onset, but it eventually diminishes the efficiency of $\widetilde{\alpha}$ for the final prevalence. This is because highly active nodes trigger the early onset but eventually suppress the final prevalence to the entire network.

It is found that optimal conditions for the temporal spread of epidemic are sensitive to connectivity patterns. However, the epidemic threshold and the final prevalence are governed by the degree distribution that is generated by temporal connectivity patterns with the heterogeneity of nodal activities $\gamma$ and the strength of memory $\beta$. The dashed lines in Fig.~\ref{fig6:heatmap} are drawn by the degree exponent $\gamma_k=(\gamma-\beta)/(1-\beta)$ of network structures. 

\section{Summary and remarks}
\label{summary}

To sum up, we revisited the effect of temporal connectivity patterns on epidemic process, in terms of the susceptible-infected-recovered (SIR) model on our modified activity-driven temporal networks (ADTNs) with memory, where we focused on how the epidemic threshold and the outbreaks are affected by temporal connectivity patterns with the activity heterogeneity and the strength of memory. The epidemic threshold was found to be directly related to the degree exponent of accumulated connectivity patterns in ADTNs, even though the networks were generated with different connectivity patterns. We also found that the initial spreading speed of epidemic is inversely proportional to the value of the epidemic threshold. Despite this, the case where the epidemic threshold is small cannot lead to a large outbreak size. Such behaviors are caused by memory and the activity heterogeneity in modified ADTNs. Memory that often establishes strong ties makes the spread of epidemic slow, which implies that the epidemic threshold increases and the outbreak size decreases as the strength of memory becomes stronger. The heterogeneity of nodal activities reduces the epidemic threshold, while the dominance of a few highly active nodes inhibits the propagation of epidemic to the entire network. 

Based on our results, we presented the optimal strategies for the spread of epidemic in temporal networks: To successful propagate an epidemic in a short period of time, the presence of highly active nodes (in time-accumulated network representation, hubs with larger degrees), is important because it may be a good strategy to focus on a few highly active nodes and support the activation of such nodes. Otherwise, to propagate the epidemic throughout the whole network, it is better to give equal opportunities to all nodes rather than to focus on a few of active one. Ultimately, we enable to maximize the efficiency of the spread using both strategies by controlling the characteristics of temporal connectivity patterns as well as network structures according to the temporal stages of the epidemic. Therefore, the optimal conditions for each stage can be suggested by considering the trade-off between the time cost and the infection rate.
 
There are still important tasks remaining, such as the numerical confirmation of the activity-based mean-field approximation with memory~\cite{Tizzani2018}, the finite-size scaling analysis of the SIR model with the comparison of that in static networks, and dynamic properties of connectivity patterns according to the time resolution of ADTNs, similar to~\cite{Kim2018}. However, these are out of our scope in this paper since we here like to focus on the validity check for our scenario that the degree distribution determines the epidemic threshold even in ADTNs, while temporal connectivity patterns control the final prevalence and the optimal condition for the spread of epidemic. 
This is related to spreading phenomena that arise from the monopoly of minority groups in the real world, for example, in the development of technology and the expansion of business. Our study might highlight the importance of the balance of activities between a small number of leader groups and other groups for the early growth and overall development. As a possible future work, it would be interesting to apply our study to real-world data or to find the optimal condition in other types of connectivity patterns, such as triadic-closure connections.

%
% The section below may be edited at your convenience to acknowledge 
% each author's contribution to the manuscript.
% You may remove it if you are a single author.
%
\section*{Authors contribution statement}
All authors designed the research and wrote the manuscript equally. Particularly, numerical simulations were performed by H.K. and analytic results with intuitive arguments were developed by M.H., and the validity check of main results were confirmed by all authors.%

\bigskip

\noindent{This research was supported by
Basic Science Research Program through the National
Research Foundation of Korea (NRF) (KR) [Grants
No. NRF-2017R1A2B3006930  (H.K., H.J.) and NRF-2017R1D1A3A03000578 (M.H.)].}

\bibliography{2019EPJB-KHJ-ref}

% BibTeX users please use
% \bibliographystyle{}
% \bibliography{}
%\begin{thebibliography}{}
%Author, \textit{Book title} (Publisher, place year) page numbers
% etc
%\end{thebibliography}

\end{document}